\documentclass[
   aps,
    pra,
    onecolumn,
    letterpaper,
    10pt,
    showpacs,
    notitlepage,
    amsmath,
    amssymb,
    floatfix
]{revtex4}

\usepackage{bm,enumerate,dcolumn,tikz,graphicx,color,amsmath,amssymb}
\usepackage{epstopdf}
\usepackage{empheq}
\usepackage{capt-of}
\usepackage{pgfplots}
\usepackage{hyperref}% add hypertext capabilities

\newcommand{\be}{\begin{equation}}
\newcommand{\ee}{\end{equation}}

\begin{document}
\title{Hyperfine-induced quadrupole moments of alkali-metal atom ground states and their implications for atomic clocks}
\author{Andrei Derevianko}
\affiliation{Department of Physics, University of Nevada, Reno, NV 89557, USA}
\date{\today}

\begin{abstract}
 Spherically-symmetric  ground states of alkali-metal atoms do not posses  electric quadrupole moments. However, the hyperfine interaction between nuclear moments and atomic electrons distorts  the spherical symmetry of electronic clouds and leads to non-vanishing atomic quadrupole moments. We evaluate these hyperfine-induced quadrupole moments using techniques of relativistic many-body theory and
 compile results for Li, Na, K, Rb, and Cs atoms. For heavy atoms we find that the hyperfine-induced quadrupole moments are strongly (two orders of magnitude) enhanced by correlation effects.
We further apply the results of the calculation to microwave atomic clocks where the coupling of
 atomic quadrupole moments  to gradients of electric fields leads to clock frequency uncertainties. 
We show that for $^{133}$Cs atomic clocks, the spatial gradients of  electric fields must be smaller than 
$30 \, \mathrm{V}/\mathrm{cm}^2$ to guarantee fractional inaccuracies below $10^{-16}$.
\end{abstract}

% pacs, the Physics and Astronomy Classification Scheme
%\pacs{42.50.Dv, 03.67.-a, 32.80.Rm, 78.67.-n}
%31.15.A-	Ab initio calculations
%32.10.Dk	Electric and magnetic moments, polarizabilities
%06.30.Ft	Time and frequency

\maketitle

\section{Introduction}
Hyperfine-interaction(HFI)-induced effects are ubiquitous in atomic physics. Examples include hyperfine quenching of otherwise E1-forbidden transitions~\cite{Gar62,Joh2011_HFIquenching,PorDer04}, differential AC Stark shifts in ground states of alkalis~\cite{DzuFlaBel10}, electron bridge  in nuclear transitions~\cite{PorFla2010}, black-body radiation shift in atomic clocks~\cite{BelSafDer06,AngDzuFla06}, ``magic'' trapping of hyperfine qubits~\cite{ChiNelOlm10}, etc. In all these cases the effects come from the HFI-mediated admixtures to atomic states. Here we focus on computing HFI-induced quadrupole moments of ground-state alkali-metal atoms.

For neutral atoms, the electric-quadrupole moment  (E2) is the leading electric multipole moment, expectation value of which does not vanish identically. Electric quadrupoles couple to gradients of electric fields and can give rise to atomic energy shifts, thereby affecting accuracy of atomic clocks~\cite{Ita00}. Because it is a second-rank tensor, the expectation value of quadrupole moment operator vanishes for the spherically-symmetric electronic ground states of alkali-metal atoms. However, the hyperfine interaction between nuclear moments and atomic electrons can distort  the spherical symmetry of electronic clouds and can lead to non-vanishing atomic quadrupole moment. Calculation of such hyperfine-induced quadrupole moments for ground-state alkali-metal atoms is the goal of this paper. We also evaluate the clock frequency shifts in microwave clocks arising from coupling of atomic quadrupole moments to E-field gradients.

\section{Formalism}
We develop the formalism in terms of the hyperfine states 
$|\gamma (IJ)FM_{F}\rangle$. Here the nuclear spin $I$ and  the total electronic angular momentum $J$ are conventionally
coupled yielding the state of the angular momentum $F$ and its projection
$M_{F}$, with $\gamma $ encapsulating all other atomic quantum numbers. When there is no ambiguity we will use a 
shorthand notation  $|\gamma FM_{F}\rangle$. For the ground $S_{1/2}$ states of alkali-metal atoms the two allowed values of 
$F$ are $F=I+1/2$ and $F=|I-1/2|$. As long as $F \ge 1$, the hyperfine state will have non-vanishing quadrupole moment.

The irreducible tensor operator of electric-quadrupole moment  is defined as
\begin{equation}
Q_{\mu}=-|e|\sum_{i}r_{i}^{2}C_{\mu}^{(2)}(\hat{{\mathbf{r}}}_{i})\,,
\label{Eq_Qten}
\end{equation}
where the summation is over atomic electrons, ${\mathbf{r}}_{i}$ is the
position vector of electron $i$, and $C_{\mu}^{(2)}(\hat{{\mathbf{r}}}_{i})$
are normalized spherical harmonics~\cite{VarMosKhe88}. The quadrupole moment ${\mathcal{Q}}$ of a  state of definite angular momentum $|\gamma F,M_{F} \rangle$ is defined conventionally as twice
the expectation value in the stretched state
\begin{align}
{\mathcal{Q}}=&2\,\langle\gamma F,M_{F}=F|\,Q_{0}\,|\gamma F,M_{F}%
=F\rangle\,.
\end{align}
This quadrupole moment is related to the reduced matrix element of the tensor,
Eq.~(\ref{Eq_Qten}) via the Wigner-Eckart theorem.

Total atomic quadrupole  may be  decomposed into the direct (no HFI) electronic-cloud contribution,  the direct nuclear contribution,  and a correction due to distortion of electronic-cloud charge distribution by the HFI,
\[
 \mathcal{Q} = \mathcal{Q}^\mathrm{elec} + \mathcal{Q}^\mathrm{nuc}  + \mathcal{Q}^\mathrm{HFI} \,.
\]
For the ground $S$ state of alkalis, the direct electronic contribution $\mathcal{Q}^\mathrm{elec}$  vanishes due to the spherical symmetry of charge distribution.
The values of nuclear moments are on the order of $10\, |e| \, \mathrm{fm}^2 \approx 4 \times 10^{-9} \, |e| a_0^2$ (see Table~\ref{Tab:nuc-params}). We show below that $\mathcal{Q}^\mathrm{HFI}$
for alkali-metal atoms ranges  from $10^{-8} \, |e| a_0^2$ (Li) to $10^{-5} \, |e| a_0^2$ (Cs), thereby the HFI-induced contribution dominates at least for heavy alkali-metal   atoms.

\begin{table}[htp]
\caption{Compilation of nuclear radii $R_\mathrm{nuc}$ and nuclear magnetic and electric quadrupole moments used in calculations. Values of $R_\mathrm{nuc}$ are from~\cite{JohSof85} and nuclear moments are from~\cite{Stone2005}. 
\label{Tab:nuc-params} }
\begin{center}%
\begin{tabular}
[c]{lllllll}\hline\hline
Isotope                                 	& $^{7}$Li 	& $^{23}$Na 	& $^{39}$K	&  $^{85}$Rb 	& $^{87}$Rb 	& $^{133}$Cs 	\\\hline
$Z$                                      	& 3 			& 11 		  	& 19 			& 37			& 37  		& 55 			\\
$I$                                       	& 3/2 		& 3/2 	  	& 3/2 		& 5/2 		& 3/2			& 7/2 		\\
$R_\mathrm{nuc}$, fm        	& 1.80 		& 2.89 	  	& 3.61 		&4.87 		& 4.87 		&  5.67 		\\
$\mu_\mathrm{nuc}/\mu_{N}$ 	& 3.256 	         & 2.218 	        & 0.3915 	         &1.353 	        & 2.752     	& 2.583	 \\
$Q_\mathrm{nuc}, |e| \,\mathrm{fm}^2$ 
                                              	& -3.7		&10.1		& 4.9		& 27.3		& 13.2 		&-0.37              \\
\hline\hline
\end{tabular}
\end{center}
\label{default}
\end{table}%

 In the
first order of perturbation theory in the hyperfine interaction, $H_{\mathrm{%
HFI}}$, the correction to the hyperfine sub-level $|\gamma (IJ)FM_{F}\rangle
$ attached to electronic state $|\gamma J\rangle $ reads
%===========================================================================
\begin{equation}
|\gamma (IJ)FM_{F}\rangle ^{\mathrm{HFI} }=\sum_{\gamma ^{\prime
}J^{\prime }}|\gamma ^{\prime }(IJ^{\prime })FM_{F}\rangle \frac{\langle
\gamma ^{\prime }(IJ^{\prime })FM_{F}|H_{\mathrm{HFI}}|\gamma
(IJ)FM_{F}\rangle }
{E\left( \gamma J\right) -E\left(\gamma ^{\prime }J^{\prime} \right) }\,,  \label{Eq:WFCorr}
\end{equation}
%===========================================================================
where $E\left( \gamma J\right) $ are the energies of atomic states. In the
above expression, we have taken into account that $H_{\mathrm{HFI}}$ is a
scalar, so the total angular momentum $F$ and its projection $M_{F}$ are
conserved. 
In particular, for the alkali-metal atoms in the $S_{1/2}$ states, the HFI admixes $D$ states which distort the spherical symmetry of the $S$ states leading to non-vanishing quadrupole moments.
Explicitly,
\begin{align*}
\langle \gamma (IJ)FM_{F}|Q|\gamma (IJ)FM_{F}\rangle^{\mathrm{HFI} } =&2 \sum_{\gamma ^{\prime
}J^{\prime }}
\langle \gamma (IJ)FM_{F}|Q|\gamma ^{\prime }(
IJ^{\prime })FM_{F}\rangle \times \\
&\frac{\langle 
\gamma ^{\prime }(IJ^{\prime })FM_{F}|H_{\mathrm{HFI}}|\gamma
(IJ)FM_{F}\rangle }
{E\left( \gamma J\right) -E\left(\gamma ^{\prime }J^{\prime} \right) }
\,.  \label{Eq:QCorr}
\end{align*}
For the ground state of alkali-metal atoms, the intermediate states are of the $D_{3/2}$ and $D_{5/2}$ character.

In general, the hyperfine coupling Hamiltonian, $H_{\mathrm{HFI}}$, may be represented as a sum over multipole nuclear moments $\mathcal{N}%
^{(k)}_\lambda$ of rank $k$ combined with the even-parity electronic coupling
operators $\mathcal{T}^{(k)}_\lambda$ of the same rank so that the total interaction
is rotationally and P-- invariant,
\[
H_{\mathrm{HFI}}=\sum_k \left( \mathcal{N}^{(k)}\cdot \mathcal{%
T}^{\left( k\right) }\right) \, .
\]
In the following we limit the discussion to the dominant contributions from
the nuclear magnetic-dipole ($k=1$) and electric-quadrupole ($k=2$) moments.
The conventionally defined nuclear moments are related to the tensors
$\mathcal{N}^{(k)}_\lambda$ as
$\mu \equiv \langle I M_I=I|  \mathcal{N}^{(1)}_0 |I M_I=I \rangle$ and
$\mathcal{Q}^\mathrm{nuc} \equiv 2 \langle I M_I=I|  \mathcal{N}^{(2)}_0 |I M_I=I \rangle$. These nuclear moments are compiled in Table~\ref{Tab:nuc-params}.
Explicit expressions for electronic tensors $\mathcal{T}^{(k)}_\lambda$ can be found in Ref.~\cite{Joh07book}.

Using angular momenta algebra,
the matrix element of the hyperfine interaction
in Eq.~(\ref{Eq:WFCorr}) may be simplified to
\begin{align}
&\langle \gamma ^{\prime }\left( IJ^{\prime }\right) F^{\prime}M^{\prime}_{F}|H_{\mathrm{HFI}%
}|\gamma \left( IJ\right) FM_{F}\rangle =
\delta _{FF^{\prime}}\delta _{M_{F}M_{F}^{\prime }}\times  \\ \nonumber
&(-1)^{I+J^{\prime }+F} \sum_k \langle I\ ||\mathcal{N}%
^{(k)}||\ I\rangle \langle \gamma^{\prime }J^{\prime }||\mathcal{T}^{\left(
k\right) }||\gamma J\rangle \left\{
\begin{tabular}{lll}
$I$ & $I$ & $k$ \\
$J$ & $J^{\prime }$ & $F$%
\end{tabular}
\right\} .
\end{align}
Reduced matrix elements  of $Q$ between the hyperfine  states read
\begin{align*}
\langle  \gamma (IJ)F ||Q|| \gamma ^{\prime }(IJ^{\prime })F \rangle &
=(-1)^{J+I+F}\; (2F+1) 
\left\{
\begin{array}
[c]{ccc}%
J & F & I\\
F & J' & 2
\end{array}
\right\}  
 \langle \gamma J  ||Q|| \gamma ^{\prime }J^{\prime} \rangle \,.
\end{align*}

Thereby the conventional quadrupole moment of the $|\gamma (IJ)F\rangle$ state is
\begin{align}
{\mathcal{Q}}_F^{\mathrm{HFI}}=
2 
\left(
\begin{array}
[c]{ccc}%
F & 2 & F\\
-F & 0 & F
\end{array}
\right)
\langle \gamma (IJ)F ||Q||\gamma (IJ)F\rangle^{\mathrm{(HFI)}}
\end{align}
with
\begin{align*}
\langle \gamma (IJ)F ||Q||\gamma (IJ)F\rangle^{\mathrm{(HFI)}}  =&2 (2F+1) (-1)^{2F+2I} \; \times\\
&\sum_{kJ'}
\left\{
\begin{tabular}{lll}
$I$ & $I$ & $k$ \\
$J$ & $J^{\prime }$ & $F$%
\end{tabular}
\right\}
\left\{
\begin{array}
[c]{ccc}%
J & F & I\\
F & J' & 2
\end{array}
\right\} 
S_{J'}^{(k)}
\end{align*}
expressed in terms of purely electronic sums
\begin{align}
S_{J'}^{(k)} = (-1)^{J+J^{\prime }} \langle I\ ||\mathcal{N}^{(k)}||\ I\rangle  \sum_{\gamma ^{\prime
}}  
\frac{\langle \gamma J||Q||\gamma ^{\prime }J^{\prime }\rangle
\langle \gamma^{\prime }J^{\prime }||\mathcal{T}^{\left(
k\right) }||\gamma J\rangle }
{E\left( \gamma J\right) -E\left(\gamma ^{\prime }J^{\prime} \right) }
\,.  \label{Eq:SkJp}
\end{align}
Both magnetic-dipole and electric-quadrupole HFI contribute to the electronic quadrupole moment of the $S_{1/2}$ ground state. For the $D_{3/2}$ channel $k=1,2$ and for the $D_{5/2}$ channel $k=2$ (only the nuclear quadrupole HFI contribution).

If the magnetic-dipole HFI channel ($k=1$) dominates, the ratio of HFI-induced quadrupole moments for the two hyperfine components is given by ($I \ge 3/2$)
\begin{equation}
 \frac{ {\mathcal{Q}}^{\mathrm{HFI}}_{F=I+1/2} } { {\mathcal{Q}}^{\mathrm{HFI}}_{F=I-1/2} } =    - \frac{ (2I) (2I+1)}{ (2I-1)(2I-2)} \label{Eq:Q-Ratio} \,.
\end{equation}

\section{Relativistic many-body calculations and results}
We employ several relativistic many-body methods of atomic structure to evaluate the HFI-induced quadrupole moments.  We carry out computations in several approximations of increasing complexity. The lowest order approximation in the electron-electron interaction is the mean-field Dirac-Hartree-Fock (DHF) method. We used the so-called frozen-core approximation, where the DHF potential was computed self-consistently for core orbitals and the valence orbitals were determined in the resulting  $V^{N-1}$ DHF potential. The summation over intermediate states in Eq.~(\ref{Eq:SkJp}) was carried out using the dual-kinetic balance (DKB) B-spline technique~\cite{BelDer08} which guarantees numerical quality of orbitals both near and far away from the nucleus. The DKB basis orbitals were generated in the $V^{N-1}$ DHF potential.

The required relativistic one-particle reduced matrix element of quadrupole moment operator in atomic units is given by
\begin{equation}
\langle\phi_{i}||Q||\phi_{j}\rangle=-\langle\kappa_{i}||C^{(2)}||\kappa
_{j}\rangle\,\int_{0}^{\infty}r^{2}[P_{i}(r)P_{j}(r)+Q_{i}(r)Q_{j}(r)]dr\,,
\label{Eq_melQ}%
\end{equation}
where $P$ and $Q$ are the large and small radial components of Dirac bi-spinor wave function and $\kappa$ are relativistic angular quantum numbers~\cite{Joh07book}. Minus sign accounts for the negative charge of the electron. Expressions for reduced matrix elements of electronic HFI tensors $\mathcal{T}^{(k)}_\lambda$ 
between Dirac bi-spinors 
can be found in Ref.~\cite{Joh07book}.

 An addition of core-polarization effects by the valence electron leads to the Brueckner-orbital (BO) approximation. In this approximation the core-polarization potential, computed as the second-order self-energy correction, was added to the DHF potential and the resulting Hamiltonian was diagonalized in the DKB basis resulting in a complete set of Brueckner orbitals. This numerically complete BO set was used in evaluation of the sums over intermediate states~(\ref{Eq:SkJp}).
 
 Further in addition to the DHF and BO methods we also included an infinite chain of random-phase approximation (RPA) diagrams. 
 The RPA diagrams account for the self-consistent  screening of the HFI and quadrupole operators by core electrons.
 The two resulting approximations are DHF+RPA and BO+RPA methods distinguishing between the two distinct basis sets (DHF and BO) used in the evaluation of RPA diagrams. The most complete approach which includes both the core polarization and the core  screening effects is the BO+RPA method and our final (recommended) values are the BO+RPA values.
  
The quality of the enumerated approximations has been studied extensively in the literature (see, e.g., Refs.~\cite{JohLiuSap96,DzuFlaSus84}). For example, the experimental values of ground-state hyperfine structure constants $A$ are reproduced at the level of a few per cent for heavy alkali-metal atoms.
Based on this prior literature, we expect  the BO+RPA approximation 
accuracy to be better than 10\% for heavy atoms and a few per cent for light atoms.
 While more sophisticated and more accurate all-order methods do exist (for example, the hyperfine constant $A_{6s}$ for Cs is reproduced with the accuracy of a few 0.1\%  in Ref.~\cite{PorBelDer10}), the BO+RPA approximation should be adequate for the goals of this paper. 
 
\begin{table}[htp]
\caption{Hyperfine-induced quadrupole moments of the $F=3$ and $F=4$ hyperfine states of the $6S_{1/2}$ ground state of $^{133}$Cs in various many-body approximations. Values are given in $|e| a_0^2$. \label{Tab:Q-values}}
\begin{center}
\begin{tabular}{lll}
\hline\hline
Approximation & $\mathcal{Q}^{\mathrm{HFI}}_{F=3}$ & $\mathcal{Q}^{\mathrm{HFI}}_{F=4}$   \\[0.5ex]
\hline
DHF                 &  $-3.5 \times 10^{-8}$    &   $+6.6 \times 10^{-8}$              \\[0.5ex]
BO                   &  $-2.7 \times 10^{-8}$     &    $+5.3 \times 10^{-8}$              \\[0.5ex]
DHF+RPA        &  $+6.3  \times 10^{ -6}$ &    $-1.2 \times 10^{-5}$              \\[0.5ex]
BO+RPA [final] & $+8.4  \times 10^{ -6}$  &    $-1.6 \times 10^{-5}$             \\
\hline\hline
\end{tabular}
\end{center}
\label{default}
\end{table}%

\begin{figure}[h!tb]
	\begin{center}
		\includegraphics[width=0.45\textwidth]{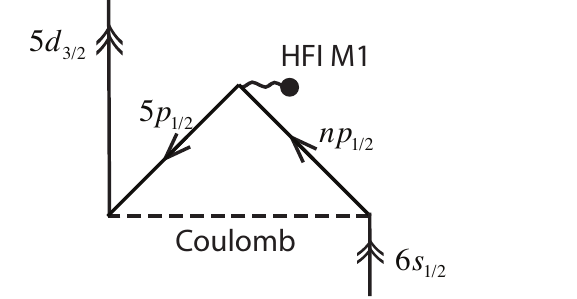}
  \end{center}
\caption{  Second-order RPA contribution to the matrix element of magnetic-dipole hyperfine interaction between the $6s_{1/2}$ and $5d_{3/2}$ valence states in Cs. Such diagrams   
lead to the substantial correlation enhancement of the HFI-induced quadrupole moments.
	}
	\label{Fig:RPAdom}
\end{figure}

%We employed several many-body approximations (see Table~\ref{Tab:Q-values}).  
%We use the following nuclear parameters for $^{133}$Cs: $R_m = 5.6748 \, \mathrm{fm}$, $g_I =( \mu/\mu_n)/I =  0.73789$, $Q_\mathrm{nuc} = -0.371 \, |e| \, \mathrm{fm}^2$.
As an illustrative example, we consider the evaluation of the HFI-induced quadrupole moment of $^{133}$Cs (see Table~\ref{Tab:Q-values}).  We observe that  core polarization (BO) changes the lowest-order (DHF) values 
by about 30\%. The core-screening effect (RPA) is much more significant, as it not only changes the sign of the computed quadrupole moment, but increases it in absolute value by two orders of magnitude. The reason for this strong enhancement (supported by our numerical second-order RPA analysis) is shown in Fig.~\ref{Fig:RPAdom}.
Indeed when the sums (\ref{Eq:SkJp}) are computed in the lowest (DHF or BO) order, 
the matrix elements of the HFI interaction are between the $d_{3/2}$ and $6s_{1/2}$ states. By contrast, the RPA diagram, Fig.~\ref{Fig:RPAdom}, involves the matrix element of the HFI interaction evaluated between $p_{1/2}$ states. This matrix element is much larger than the $d_{3/2}- 6s_{1/2}$ matrix element due to suppressed overlap of $d$-orbitals with the nuclear region.

%There is a strong RPA-correlation enhancement of the magnetic-HFI $6s-5d_{3/2}$ matrix element that increases the value of $\mathcal{Q}$ by two orders of magnitude and changes it's sign.
%Plausibly the RPA diagram $6s_{1/2} - 5d_{3/2} --> L=1--> 5s_{1/2}-AM1-6s_{1/2}$ is the reason b/c of a very large  M1 hyperfine interaction matrix element between $s$-states.

The dominant contribution to the HFI-induced quadrupole moment of $^{133}$Cs  comes from the magnetic-dipole interaction with the nucleus ($d_{3/2}$ channel).   The ratio of computed BO+RPA quadrupole moments for Cs $F=4$ and $F=3$ states, $-1.87$, is in a good agreement with the analytical ratio~(\ref{Eq:Q-Ratio}) of $-28/15$ due to negligibly small contributions from the quadrupole HFI interaction ($k=2$).

Finally we compile  the computed HFI-induced quadrupole moments for alkali-metal atoms in Table~\ref{Tab:Q-HFI-all}. The employed nuclear parameters are listed in Table~\ref{Tab:nuc-params}. All the listed values of $\mathcal{Q}^{\mathrm{HFI}}$ in Table~\ref{Tab:Q-HFI-all} are computed in the BO+RPA approximation. In all atoms except Li, the RPA corrections lead to the sign reversal of  $\mathcal{Q}^{\mathrm{HFI}}$ DHF (or BO) values.

\begin{table}[htp]
\caption{Hyperfine-induced quadrupole moments of the hyperfine states of the $S_{1/2}$ ground states of alkali-metal atoms in the BO+RPA approximation. Values are given in $|e| a_0^2$. 
Notation $x[y]$ stands for $x \times 10^y$.
\label{Tab:Q-HFI-all} }
\begin{center}%
\begin{tabular}
[c]{lrrrrrr}\hline\hline
Isotope                                 	& $^{7}$Li 	& $^{23}$Na 	& $^{39}$K	&  $^{85}$Rb 	& $^{87}$Rb 	& $^{133}$Cs 	\\\hline
$I$                                       	& 3/2 		& 3/2 	  	& 3/2 		& 5/2 		& 3/2			& 7/2 		\\
$\mathcal{Q}^{\mathrm{HFI}}_{F=I+1/2} $
                                              &   6.0[-8]               &  -5.8[-8]        &     -4.6[-7]       &   -2.3[-6]         &  -4.7[-6]          &  -1.6[-5]                \\
                                             
$\mathcal{Q}^{\mathrm{HFI}}_{F=I-1/2}$
                                              &    -1.0[-8]            &   1.1[-8]          &      7.8[-8]      &    9.4[-7]        &  7.9[-7]            &    8.4[-6]             \\

%$g_{I}$ 				  	& 2.17098 	& 1.47844   	& 0.61005 	& 0.541341 	& 1.83455		& 0.737975 	 \\
\hline\hline
\end{tabular}
\end{center}
\label{default}
\end{table}%

\section{Implications for atomic clocks}

 Atomic quadrupole moments couple to gradients of electric field. The interaction perturbs atomic energy levels, leading to clock frequency uncertainties. The relevant clock shifts for singly-charged ion clocks are sizable and considerable efforts have been devoted to mitigating this effect~\cite{Ita00,RosHumSch08}. Below we evaluate the size of such an effect for microwave clocks based on hyperfine transitions in alkali-metal atoms.

  The interaction  of atomic quadrupole moments with gradients of electric field can be represented as
 \begin{equation}
 {\hat V_Q} =  - \frac{1}{{\sqrt 6 }}\left( {{\hat{Q}^{\left( 2 \right)}} \cdot {{\left( {\nabla \otimes \mathcal{E}} \right)}^{\left( 2 \right)}}} \right) = - \frac{1}{{\sqrt 6 }} \sum\limits_{\lambda  =  - 2}^2 {{{\left( { - 1} \right)}^\lambda }{Q^{\left( 2 \right)}_{-\lambda} }} {\left( {\nabla \otimes  \mathcal{E
}} \right)_\lambda^{\left( 2 \right)}} \,.
\label{Eq:VQ}
 \end{equation}
 Here we explicitly emphasized that the quadrupole moment operator~(\ref{Eq_Qten})  is a rank 2 tensor. The second-rank tensor of the E-field gradient is obtained  by conventionally coupling  the gradient and E-field vectors 
 \begin{equation}
 \left( {\nabla  \otimes \mathcal{E}} \right)_\lambda ^{\left( 2 \right)} = \sum\limits_{q,q'} {C_{1q,1q'}^{2\lambda }} {\nabla _q}{\mathcal{E}_{q'}} \, .
 \label{Eq:gradEtensor}
 \end{equation}
 This tensor is expressed in terms of Clebsch-Gordan coefficients and spherical components of the gradient and E-field~\cite{VarMosKhe88}. The E-field derivatives are evaluated at the  location of the atom. 

The interaction   $\hat V_Q$ can lead to transitions between different magnetic sub-levels and also to energy shifts. The former effect can be neglected as long as the Zeeman splittings are much larger than the off-diagonal matrix elements of $\hat V_Q$ and latter can lead to clock frequency shift which is the main focus of this section. 

The lowest-order energy shift of an atomic level $|\gamma FM_{F}\rangle$ reads
\be
\delta E_{\gamma F M_F} =  - \frac{1}{2}\left\langle {{Q_0}} \right\rangle_{\gamma F M_F} \frac{{\partial {\mathcal{E}_z}}}{{\partial z}} \,.
\ee
While arriving at this expression from Eqs.(\ref{Eq:VQ},\ref{Eq:gradEtensor})  we used the fact that $\nabla \cdot  \mathbf{\mathcal{E}} = \sum_{\lambda}
(-1)^{\lambda} \nabla_{-\lambda} \mathcal{E}_\lambda=0$.
In a spatially uniform field gradient along the length of an atomic fountain, the clock frequency shift $\delta \nu_Q$ reads
\be
h \delta \nu_Q = -\frac{1}{2} \Delta Q   \frac{{\partial {\mathcal{E}_z}}}{{\partial z}} \,,
\ee
where the differential quadrupole moment is $\Delta Q = \left\langle {{Q_0}} \right\rangle_{\gamma F' M'_F} - \left\langle {{Q_0}} \right\rangle_{\gamma F M_F}$ , with $F' M'_F$ and  $F M_F$ being the upper and  the lower clock levels. 
For the B-field-insensitive Zeeman sub-levels, $M_F=M'_F =0$, the matrix elements can be expressed in terms of the computed quadrupole moments as ($F \ge1$)
\be
\left\langle {{Q_0}} \right\rangle_{\gamma F M_F=0} =- \frac{1}{2} \frac{F+1}{2F-1} \mathcal{Q}_F .
\ee
When further the relation~(\ref{Eq:Q-Ratio}) holds, the differential quadrupole moment between the $F'=I+1/2,M_F'=0$ and $F=I-1/2,M_F=0$  hyperfine levels can be simplified further ($I\ge 3/2$)
\be
\Delta Q = \frac{ (2I+1)^2}{2(2I-1)(2I-2) } \mathcal{Q}_{F=I-1/2} \, .
\ee
For example for the clock transition in $^{133}\mathrm{Cs}$, $I=7/2$, $F'=4$ and $F=3$, and we arrive at $\Delta Q (^{133}\mathrm{Cs} )=9.1 \times 10^{-6} |e| a_0^2$.
 We further obtain a practical formula for fractional clock inaccuracy
\be
 \frac{\delta \nu_Q}{\nu_\mathrm{clock}} \left(^{133}\mathrm{Cs}\right) = - 3.4 \times 10^{-18}  \times \left(  \frac{ \partial\mathcal{E}_z /{\partial z}}{
 \mathrm{V}/\mathrm{cm}^2 }\right)\, .
\ee
In particular we conclude that to keep the Cs clock accurate to the $10^{-16}$ level which is the current goal~\cite{HaeDonLev2014,LiGibSzy2011}, E-field gradients must be smaller than $30 \,  \mathrm{V}/\mathrm{cm}^2$.

I would like to thank L. Hunter, D. Budker and K. Gibble for motivating  discussions. This work was supported in part by the US National Science Foundation.

%\bibliographystyle{unsrt}
%\bibliography{library}

\begin{thebibliography}{20}
\expandafter\ifx\csname natexlab\endcsname\relax\def\natexlab#1{#1}\fi
\expandafter\ifx\csname bibnamefont\endcsname\relax
  \def\bibnamefont#1{#1}\fi
\expandafter\ifx\csname bibfnamefont\endcsname\relax
  \def\bibfnamefont#1{#1}\fi
\expandafter\ifx\csname citenamefont\endcsname\relax
  \def\citenamefont#1{#1}\fi
\expandafter\ifx\csname url\endcsname\relax
  \def\url#1{\texttt{#1}}\fi
\expandafter\ifx\csname urlprefix\endcsname\relax\def\urlprefix{URL }\fi
\providecommand{\bibinfo}[2]{#2}
\providecommand{\eprint}[2][]{\url{#2}}

\bibitem[{\citenamefont{Garstang}(1962)}]{Gar62}
\bibinfo{author}{\bibfnamefont{R.~G.} \bibnamefont{Garstang}},
  \bibinfo{journal}{J. Opt. Soc. Am.} \textbf{\bibinfo{volume}{52}},
  \bibinfo{pages}{845} (\bibinfo{year}{1962}).

\bibitem[{\citenamefont{Johnson}(2011)}]{Joh2011_HFIquenching}
\bibinfo{author}{\bibfnamefont{W.~R.} \bibnamefont{Johnson}},
  \bibinfo{journal}{Can. J. Phys.} \textbf{\bibinfo{volume}{89}},
  \bibinfo{pages}{429} (\bibinfo{year}{2011}),
  \urlprefix\url{http://dx.doi.org/10.1139/p11-018}.

\bibitem[{\citenamefont{Porsev and Derevianko}(2004)}]{PorDer04}
\bibinfo{author}{\bibfnamefont{S.~G.} \bibnamefont{Porsev}} \bibnamefont{and}
  \bibinfo{author}{\bibfnamefont{A.}~\bibnamefont{Derevianko}},
  \bibinfo{journal}{Phys. Rev. A} \textbf{\bibinfo{volume}{69}},
  \bibinfo{pages}{42506} (\bibinfo{year}{2004}), ISSN
  \bibinfo{issn}{1050-2947}.

\bibitem[{\citenamefont{Dzuba et~al.}(2010)\citenamefont{Dzuba, Flambaum,
  Beloy, and Derevianko}}]{DzuFlaBel10}
\bibinfo{author}{\bibfnamefont{V.~A.} \bibnamefont{Dzuba}},
  \bibinfo{author}{\bibfnamefont{V.~V.} \bibnamefont{Flambaum}},
  \bibinfo{author}{\bibfnamefont{K.}~\bibnamefont{Beloy}}, \bibnamefont{and}
  \bibinfo{author}{\bibfnamefont{A.}~\bibnamefont{Derevianko}},
  \bibinfo{journal}{Phys.\ Rev.\ A} \textbf{\bibinfo{volume}{82}},
  \bibinfo{pages}{062513} (\bibinfo{year}{2010}), ISSN
  \bibinfo{issn}{1050-2947},
  \urlprefix\url{http://link.aps.org/doi/10.1103/PhysRevA.82.062513}.

\bibitem[{\citenamefont{Porsev and Flambaum}(2010)}]{PorFla2010}
\bibinfo{author}{\bibfnamefont{S.~G.} \bibnamefont{Porsev}} \bibnamefont{and}
  \bibinfo{author}{\bibfnamefont{V.~V.} \bibnamefont{Flambaum}},
  \bibinfo{journal}{Phys. Rev. A} \textbf{\bibinfo{volume}{81}},
  \bibinfo{pages}{032504} (\bibinfo{year}{2010}), ISSN
  \bibinfo{issn}{1050-2947},
  \urlprefix\url{http://link.aps.org/doi/10.1103/PhysRevA.81.032504}.

\bibitem[{\citenamefont{Beloy et~al.}(2006)\citenamefont{Beloy, Safronova, and
  Derevianko}}]{BelSafDer06}
\bibinfo{author}{\bibfnamefont{K.}~\bibnamefont{Beloy}},
  \bibinfo{author}{\bibfnamefont{U.~I.} \bibnamefont{Safronova}},
  \bibnamefont{and}
  \bibinfo{author}{\bibfnamefont{A.}~\bibnamefont{Derevianko}},
  \bibinfo{journal}{Phys. Rev. Lett.} \textbf{\bibinfo{volume}{97}},
  \bibinfo{pages}{40801} (\bibinfo{year}{2006}), ISSN
  \bibinfo{issn}{0031-9007}.

\bibitem[{\citenamefont{Angstmann et~al.}(2006)\citenamefont{Angstmann, Dzuba,
  and Flambaum}}]{AngDzuFla06}
\bibinfo{author}{\bibfnamefont{E.~J.} \bibnamefont{Angstmann}},
  \bibinfo{author}{\bibfnamefont{V.~A.} \bibnamefont{Dzuba}}, \bibnamefont{and}
  \bibinfo{author}{\bibfnamefont{V.~V.} \bibnamefont{Flambaum}},
  \bibinfo{journal}{Phys. Rev. Lett.} \textbf{\bibinfo{volume}{97}},
  \bibinfo{pages}{40802} (\bibinfo{year}{2006}),
  \urlprefix\url{http://link.aps.org/abstract/PRL/v97/e040802}.

\bibitem[{\citenamefont{Chicireanu et~al.}(2011)\citenamefont{Chicireanu,
  Nelson, Olmschenk, Lundblad, Derevianko, and Porto}}]{ChiNelOlm10}
\bibinfo{author}{\bibfnamefont{R.}~\bibnamefont{Chicireanu}},
  \bibinfo{author}{\bibfnamefont{K.}~\bibnamefont{Nelson}},
  \bibinfo{author}{\bibfnamefont{S.}~\bibnamefont{Olmschenk}},
  \bibinfo{author}{\bibfnamefont{N.}~\bibnamefont{Lundblad}},
  \bibinfo{author}{\bibfnamefont{A.}~\bibnamefont{Derevianko}},
  \bibnamefont{and} \bibinfo{author}{\bibfnamefont{J.}~\bibnamefont{Porto}},
  \bibinfo{journal}{Phys. Rev. Lett.} \textbf{\bibinfo{volume}{106}},
  \bibinfo{pages}{063002} (\bibinfo{year}{2011}), ISSN
  \bibinfo{issn}{0031-9007}.

\bibitem[{\citenamefont{Varshalovich et~al.}(1988)\citenamefont{Varshalovich,
  Moskalev, and Khersonskii}}]{VarMosKhe88}
\bibinfo{author}{\bibfnamefont{D.~A.} \bibnamefont{Varshalovich}},
  \bibinfo{author}{\bibfnamefont{A.~N.} \bibnamefont{Moskalev}},
  \bibnamefont{and} \bibinfo{author}{\bibfnamefont{V.~K.}
  \bibnamefont{Khersonskii}}, \emph{\bibinfo{title}{{Quantum Theory of Angular
  Momentum}}} (\bibinfo{publisher}{World Scientific},
  \bibinfo{address}{Singapore}, \bibinfo{year}{1988}).

\bibitem[{\citenamefont{Johnson and Soff}(1985)}]{JohSof85}
\bibinfo{author}{\bibfnamefont{W.~R.} \bibnamefont{Johnson}} \bibnamefont{and}
  \bibinfo{author}{\bibfnamefont{G.}~\bibnamefont{Soff}}, \bibinfo{journal}{At.
  Data Nucl. Data Tables} \textbf{\bibinfo{volume}{33}}, \bibinfo{pages}{405}
  (\bibinfo{year}{1985}).

\bibitem[{\citenamefont{Stone}(2005)}]{Stone2005}
\bibinfo{author}{\bibfnamefont{N.}~\bibnamefont{Stone}}, \bibinfo{journal}{At.
  Data Nucl. Data Tables} \textbf{\bibinfo{volume}{90}}, \bibinfo{pages}{75}
  (\bibinfo{year}{2005}).

\bibitem[{\citenamefont{Johnson}(2007)}]{Joh07book}
\bibinfo{author}{\bibfnamefont{W.~R.} \bibnamefont{Johnson}},
  \emph{\bibinfo{title}{{Atomic Structure Theory: Lectures on Atomic Physics}}}
  (\bibinfo{publisher}{Springer}, \bibinfo{address}{New York, NY},
  \bibinfo{year}{2007}).

\bibitem[{\citenamefont{Beloy and Derevianko}(2008)}]{BelDer08}
\bibinfo{author}{\bibfnamefont{K.}~\bibnamefont{Beloy}} \bibnamefont{and}
  \bibinfo{author}{\bibfnamefont{A.}~\bibnamefont{Derevianko}},
  \bibinfo{journal}{Comp. Phys. Comm.} \textbf{\bibinfo{volume}{179}},
  \bibinfo{pages}{310} (\bibinfo{year}{2008}), ISSN \bibinfo{issn}{00104655}.

\bibitem[{\citenamefont{Johnson et~al.}(1996)\citenamefont{Johnson, Liu,
  Sapirstein, and Dame}}]{JohLiuSap96}
\bibinfo{author}{\bibfnamefont{W.~R.} \bibnamefont{Johnson}},
  \bibinfo{author}{\bibfnamefont{Z.~W.} \bibnamefont{Liu}},
  \bibinfo{author}{\bibfnamefont{J.}~\bibnamefont{Sapirstein}},
  \bibnamefont{and} \bibinfo{author}{\bibfnamefont{N.}~\bibnamefont{Dame}},
  \bibinfo{journal}{At.\ Data Nucl.\ Data Tables}
  \textbf{\bibinfo{volume}{64}}, \bibinfo{pages}{279} (\bibinfo{year}{1996}).

\bibitem[{\citenamefont{Dzuba et~al.}(1984)\citenamefont{Dzuba, Flambaum, and
  Sushkov}}]{DzuFlaSus84}
\bibinfo{author}{\bibfnamefont{V.~A.} \bibnamefont{Dzuba}},
  \bibinfo{author}{\bibfnamefont{V.~V.} \bibnamefont{Flambaum}},
  \bibnamefont{and} \bibinfo{author}{\bibfnamefont{O.~P.}
  \bibnamefont{Sushkov}}, \bibinfo{journal}{J. Phys. B}
  \textbf{\bibinfo{volume}{17}}, \bibinfo{pages}{1953} (\bibinfo{year}{1984}).

\bibitem[{\citenamefont{Porsev et~al.}(2010)\citenamefont{Porsev, Beloy, and
  Derevianko}}]{PorBelDer10}
\bibinfo{author}{\bibfnamefont{S.~G.} \bibnamefont{Porsev}},
  \bibinfo{author}{\bibfnamefont{K.}~\bibnamefont{Beloy}}, \bibnamefont{and}
  \bibinfo{author}{\bibfnamefont{A.}~\bibnamefont{Derevianko}},
  \bibinfo{journal}{Phys. Rev. D} \textbf{\bibinfo{volume}{82}},
  \bibinfo{pages}{36008} (\bibinfo{year}{2010}), ISSN
  \bibinfo{issn}{1550-7998},
  \urlprefix\url{http://link.aps.org/doi/10.1103/PhysRevD.82.036008}.

\bibitem[{\citenamefont{Itano}(2000)}]{Ita00}
\bibinfo{author}{\bibfnamefont{W.}~\bibnamefont{Itano}}, \bibinfo{journal}{J.
  Res. NIST} \textbf{\bibinfo{volume}{105}}, \bibinfo{pages}{829}
  (\bibinfo{year}{2000}).

\bibitem[{\citenamefont{Rosenband et~al.}(2008)\citenamefont{Rosenband, Hume,
  Schmidt, Chou, Brusch, Lorini, Oskay, Drullinger, Fortier, Stalnaker
  et~al.}}]{RosHumSch08}
\bibinfo{author}{\bibfnamefont{T.}~\bibnamefont{Rosenband}},
  \bibinfo{author}{\bibfnamefont{D.~B.} \bibnamefont{Hume}},
  \bibinfo{author}{\bibfnamefont{P.~O.} \bibnamefont{Schmidt}},
  \bibinfo{author}{\bibfnamefont{C.~W.} \bibnamefont{Chou}},
  \bibinfo{author}{\bibfnamefont{A.}~\bibnamefont{Brusch}},
  \bibinfo{author}{\bibfnamefont{L.}~\bibnamefont{Lorini}},
  \bibinfo{author}{\bibfnamefont{W.~H.} \bibnamefont{Oskay}},
  \bibinfo{author}{\bibfnamefont{R.~E.} \bibnamefont{Drullinger}},
  \bibinfo{author}{\bibfnamefont{T.~M.} \bibnamefont{Fortier}},
  \bibinfo{author}{\bibfnamefont{J.~E.} \bibnamefont{Stalnaker}},
  \bibnamefont{et~al.}, \bibinfo{journal}{Science}
  \textbf{\bibinfo{volume}{319}}, \bibinfo{pages}{1808} (\bibinfo{year}{2008}),
  ISSN \bibinfo{issn}{1095-9203},
  \urlprefix\url{http://www.ncbi.nlm.nih.gov/pubmed/18323415
  http://www.sciencemag.org/cgi/content/abstract/319/5871/1808}.

\bibitem[{\citenamefont{Heavner et~al.}(2014)\citenamefont{Heavner, Donley,
  Levi, Costanzo, Parker, Shirley, Ashby, Barlow, and
  Jefferts}}]{HaeDonLev2014}
\bibinfo{author}{\bibfnamefont{T.~P.} \bibnamefont{Heavner}},
  \bibinfo{author}{\bibfnamefont{E.~A.} \bibnamefont{Donley}},
  \bibinfo{author}{\bibfnamefont{F.}~\bibnamefont{Levi}},
  \bibinfo{author}{\bibfnamefont{G.}~\bibnamefont{Costanzo}},
  \bibinfo{author}{\bibfnamefont{T.~E.} \bibnamefont{Parker}},
  \bibinfo{author}{\bibfnamefont{J.~H.} \bibnamefont{Shirley}},
  \bibinfo{author}{\bibfnamefont{N.}~\bibnamefont{Ashby}},
  \bibinfo{author}{\bibfnamefont{S.}~\bibnamefont{Barlow}}, \bibnamefont{and}
  \bibinfo{author}{\bibfnamefont{S.~R.} \bibnamefont{Jefferts}},
  \bibinfo{journal}{Metrologia} \textbf{\bibinfo{volume}{51}},
  \bibinfo{pages}{174} (\bibinfo{year}{2014}),
  \urlprefix\url{http://stacks.iop.org/0026-1394/51/i=3/a=174}.

\bibitem[{\citenamefont{Li et~al.}(2011)\citenamefont{Li, Gibble, and
  Szymaniec}}]{LiGibSzy2011}
\bibinfo{author}{\bibfnamefont{R.}~\bibnamefont{Li}},
  \bibinfo{author}{\bibfnamefont{K.}~\bibnamefont{Gibble}}, \bibnamefont{and}
  \bibinfo{author}{\bibfnamefont{K.}~\bibnamefont{Szymaniec}},
  \bibinfo{journal}{Metrologia} \textbf{\bibinfo{volume}{48}},
  \bibinfo{pages}{283} (\bibinfo{year}{2011}),
  \urlprefix\url{http://stacks.iop.org/0026-1394/48/i=5/a=007}.

\end{thebibliography}
\end{document}